\documentclass[aps,prl,twocolumn,showpacs,amsmath,amssymb,superscriptaddress,preprintnumbers]{revtex4-1}
\usepackage[english]{babel}
\usepackage{amsmath,amssymb,amsfonts} 	
\usepackage{graphicx}
\usepackage[utf8]{inputenc}
\usepackage{bm}
\usepackage{xcolor}
\usepackage[normalem]{ulem}
\usepackage{SIunits}
\usepackage[colorlinks=true,linkcolor=blue,citecolor=blue,urlcolor=blue]{hyperref}
\usepackage{url}
\usepackage{textcomp}
\usepackage{lineno,hyperref}
\usepackage{rotating}
\usepackage{subfigure}
\usepackage{color} 
\definecolor{red}{rgb}{1.,0.,0.}
\usepackage{soul}
\usepackage{setspace}

\newcommand\THEOSMARVEL{Theory and Simulation of Materials (THEOS) and National Centre for Computational Design and Discovery of Novel Materials (MARVEL), {\'E}cole Polytechnique F{\'e}d{\'e}rale de Lausanne, 1015 Lausanne, Switzerland}
\newcommand\DTUENERGY{Department of Energy Conversion and Storage, Technical University of Denmark, DK-2800 Kgs. Lyngby, Denmark}
\newcommand\equally{These authors contributed equally to this work.}

\begin{document}

\title{Precision and efficiency in solid-state pseudopotential calculations}
\author{Gianluca Prandini}
\thanks{\equally}
\affiliation{\THEOSMARVEL}
\author{Antimo Marrazzo}
\thanks{\equally}
\affiliation{\THEOSMARVEL}
\author{Ivano E. Castelli}
\thanks{\equally}
\affiliation{\THEOSMARVEL}
\affiliation{\DTUENERGY}
\author{Nicolas Mounet}
\affiliation{\THEOSMARVEL}
\author{Nicola Marzari}
\email{nicola.marzari@epfl.ch}
\affiliation{\THEOSMARVEL}

\begin{abstract}
Despite the enormous success and popularity of density-functional theory, systematic  verification and validation studies are still limited in number and scope. Here, we propose a protocol to test publicly available pseudopotential libraries, based on several independent criteria including verification against all-electron equations of state and plane-wave convergence tests for phonon frequencies, band structure, cohesive energy and pressure. Adopting these criteria we obtain curated pseudopotential libraries (named SSSP or standard solid-state pseudopotential libraries), that we target for high-throughput materials screening (``SSSP efficiency'') and high-precision materials modelling (``SSSP precision''). This latter scores highest among open-source pseudopotential libraries available in the $\Delta$-factor test of equations of states of elemental solids.
\end{abstract}

\maketitle


\section{Introduction}
\noindent
In the last three decades, atomistic electronic-structure methods have radically changed the way we think at materials theory and simulations. The 1998 Nobel prize in Chemistry given for density-functional theory (DFT) acknowledges this shift of paradigm. The ability to reduce the complexity of the many-body quantum-mechanical problem involving interacting electrons and nuclei into more tractable forms and algorithms allowed to leverage inexpensive and exponentially growing computational power, in order to provide sufficiently precise predictions for a great number of materials. What followed has been a flourishing of methods to compute more and more complex materials properties, and most notably spectroscopies (such as Raman, IR, ARPES, EELS, NMR and more). Nowadays, the (over)confidence in first-principles methods is such that they are routinely used to help interpreting experiments and guide the discovery and understanding of novel materials. In particular, systematic DFT-based computational materials screening is a fast-growing field of research, as reflected by the creation, in the last few years, of several research centres worldwide on computational materials discovery. Up to now, the most visible output has been the creation of large databases of materials properties obtained from first principles, to be compared with or to augment experimental databases such as the Pauling file~\cite{PaulingFile}, COD~\cite{cod_webpage} or ICSD~\cite{icsd_webpage}.
Even from a purely financial perspective, the personnel cost for plane-wave studies is of the order of 1 billion US\$ per year~\cite{Note1},
in purchasing power parity (PPP) terms, matched by substantial hardware usage. It is ever so more remarkable that in spite all of this, the efforts of verification of the precision of the underlying pseudopotentials (PSPs) or projector augmented-wave (PAW) approximations have been minimal. Only in 2016 a multi-group effort was able to establish a baseline in the calculations of the equations of state of elemental crystals~\cite{Lejaeghere2016}. In addition to the issue of precision, or of verification (i.e. insuring that the removal of the core electrons from the calculations performs with the required tolerance), the issue of performance looms large - a softer, smoother PSP will allow straightforwardly faster calculations, both because the basis set is decreased as the computational cost scales with the square of the basis size, and because the minimization or iterative approaches can become more efficient or better pre-conditioned with a smaller basis.

Here, we introduce a PSP testing protocol based on extensive DFT and density-functional perturbation theory (DFPT) calculations of elemental solids, and select the optimal PSP for 85 elements of the periodic table.  Our protocol, named SSSP (standard solid-state pseudopotential) testing protocol, is made of a verification part, based on the $\Delta-$factor (i.e. the difference between all-electron and PSP equations of state)~\cite{Lejaeghere2014,Lejaeghere2016}, and an extensive performance-oriented part based on plane-wave convergence tests for phonon frequencies, band structures, cohesive energies and stress tensors. We underline here that the SSSP testing protocol is a protocol based on verification and not on validation (following the nomenclature of Ref.~\cite{Lejaeghere2016}). Indeed our primary goal is to test the precision (verification) of the PSPs and thus to furnish, through our testing protocol, PSP libraries that give results as close as possible to the ``exact'' theoretical results for the PBE functional, as they would be obtained through a perfectly converged all-electron calculation. We do not perform any comparison with respect to experimental results, i.e. we do not test the accuracy (validation) of the PSPs.\\
We consider up to eight (depending on the element) publicly available PSP libraries for the PBE functional~\cite{Perdew1996} and test them with the PWscf and Phonon codes of the Quantum ESPRESSO (QE) distribution~\cite{Giannozzi2009} in an automated fashion within the framework of the AiiDA~\cite{Pizzi2016} infrastructure for reproducible computational science.
AiiDA also allows straightforward dissemination of results through the Materials Cloud web platform~\cite{materialscloud_webpage}, a cloud service designed to enable data sharing in computational materials science.

In this paper, first we describe the test set of physically relevant quantities defined for our SSSP testing protocol and what are the corresponding selection criteria used in order to select the best PSPs. Then, we discuss the need for testing several properties, showing how different PSPs may yield similar results and convergence behaviours for one property but different ones for another property. 
Finally, we propose two optimal PSP libraries chosen according to the SSSP testing protocol and criteria.\\

\section{Results}

\subsection{SSSP testing protocol}
\noindent
In this work we investigate the precision and performance of several PSPs libraries available for the QE distribution~\cite{Giannozzi2009}. 
QE is an integrated suite of open-source codes for electronic-structure calculations based on DFT which uses plane-waves as basis set and PSPs to represent the electrostatic electron-ion interactions. Nowadays QE is one of the most popular DFT codes adopted by researchers for the first-principles study of materials properties, with more than 2000 citations last year, according to Google Scholar\texttrademark. \\
All the tested PSP libraries are based on the generalized gradient approximation (GGA) for the exchange-correlation functional of Perdew, Burke and Ernzerhof (PBE)~\cite{Perdew1996} and they include the three main pseudization approaches~\cite{Note2}:
norm-conserving (NC)~\cite{Hamann1979}, ultrasoft (US)~\cite{Vanderbilt1990} and projector-augmented wave (PAW)~\cite{Blochl1994}. In particular, we investigate three PAW PSP libraries (pslibrary.0.3.1~\cite{Kucukbenli2014}, pslibrary.1.0.0 high accuracy~\cite{DalCorso2014} and the library proposed by Topsakal and Wentzcovitch for the rare-earth elements~\cite{Topsakal2014}), five US libraries (GBRV (versions 1.2, 1.4 and 1.5)~\cite{Garrity2014}, pslibrary.0.3.1~\cite{Kucukbenli2014}, and pslibrary.1.0.0 high accuracy~\cite{DalCorso2014}) and two NC libraries (SG15, versions 1.0 and 1.1~\cite{Schlipf2015}). 
For a few selected elements, i.e. N, O, F and Hf, in which all the PSP libraries above perform less well in the $\Delta$-factor test we also consider the recent NC Pseudo Dojo library~\cite{Dojo2017} and, only for N, our own set (called THEOS) of US PSPs. 
Besides, for the elements from H to Ne in the periodic table, we test the NC library proposed by Willand \textit{et al.}~\cite{Willand2013} which is tailored for systems made of light elements (see Table~\ref{pseudo_libraries} for a list of all the PSP libraries tested). \\

\begin{figure*}[!hbtp]
\centering
\includegraphics[scale=0.75]{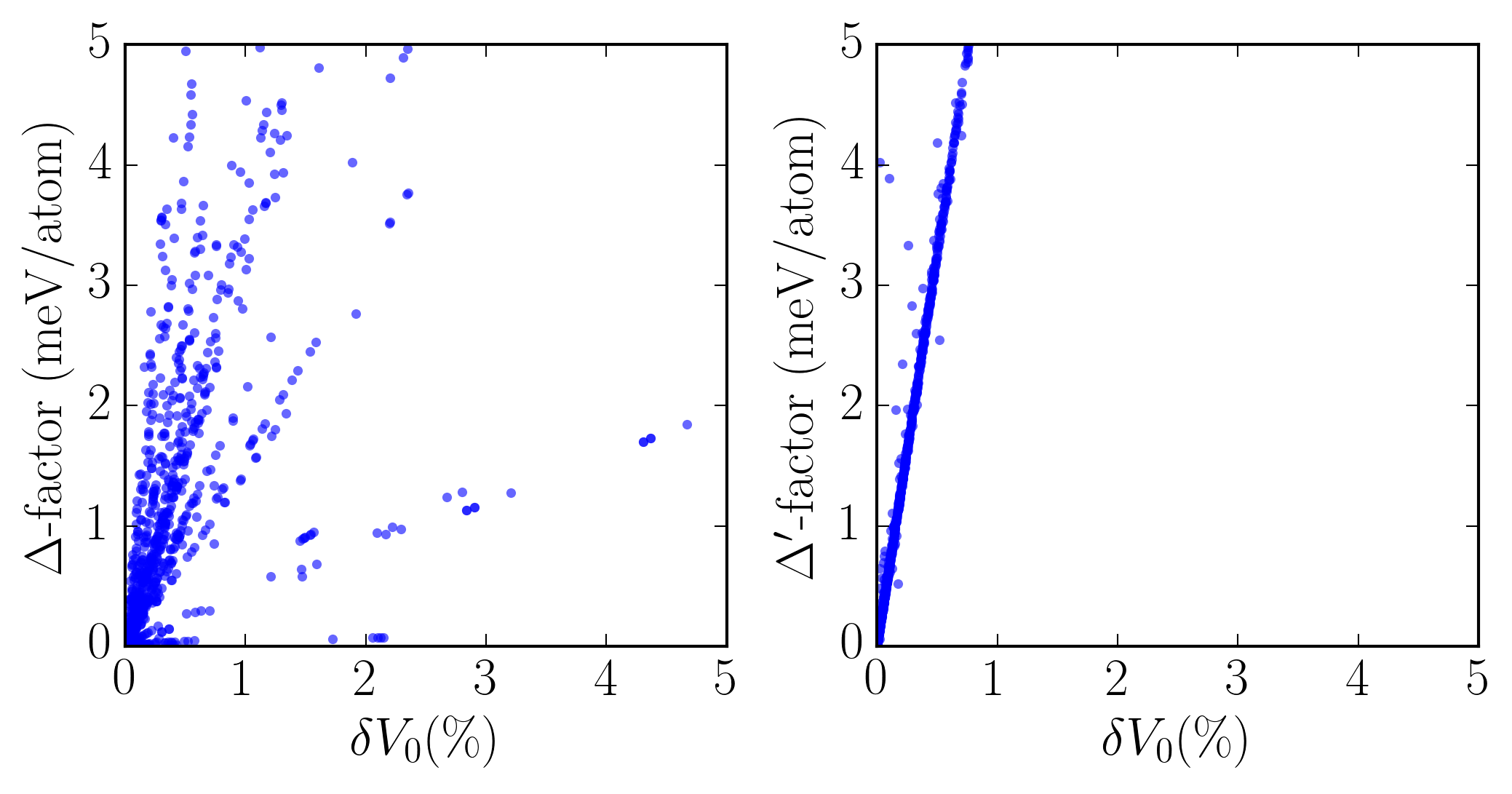}
\caption{Correlation of the $\Delta$-factor (left panel) and the $\Delta'$-factor (right panel) with the difference in equilibrium volume, $\delta V_0$, estimated from the Birch-Murnaghan fit of the equations of state. The data points are obtained from the comparison between pairs of PSPs for the 85 elemental crystals tested in this work and are the results of calculations performed at $E_c^{\text{ref}}=200 \text{ Ry}$. While the $\Delta'$-factor is strongly correlated with $\delta V_0$, the $\Delta$-factor is instead scattered along a wide range of slopes that correspond to different values of the bulk modulus $B_0$.}
\label{delta-delta1_volume}
\end{figure*}

\paragraph{Equation of state.}
In order to assess the precision of PSPs, we compute the $\Delta$-factor, i.e. the integral of the difference between the equations of state calculated with PSP simulations and with reference all-electron results. For this purpose we use the protocol introduced in 2014 by Lejaeghere \textit{et al.}~\cite{Lejaeghere2014}. This protocol was recently exploited to compare 15 different DFT codes, including both all-electron and PSP codes, in order to verify the reproducibility of the PBE equations of state of elemental crystals across different methods and implementations~\cite{Lejaeghere2016}. \\
The protocol consists in calculating the energy-versus-volume at 7 equidistant points centred around the reference equilibrium volume and then performing a Birch-Murnaghan fit. From the parameters of the fit some important physical quantities related to the structural and elastic properties of the system are extracted: the equilibrium volume $V_0$, the bulk modulus $B_0$ and the first derivative of the bulk modulus $B_1$. The $\Delta$-factor, that is reported in units of meV/atom, gives an overall estimate of the discrepancy between PSPs and all-electron results in terms of these structural properties.\\ 
However, as originally noticed by F. Jollet \textit{et al.}~\cite{Jollet2014}, the $\Delta$-factor is a stiffness dependent quantity, being proportional to $B_0$. Indeed, very soft materials, as for example the noble-gas solids, are generally associated with small values of the $\Delta$-factor, even for significant volume differences. On the other hand for very hard materials the opposite situation occurs, i.e. small volume differences give rise to large values of the $\Delta$-factor. To solve this problem the alternative $\Delta'$-factor was introduced, which corresponds to a $\Delta$-factor ``renormalized'' to reference values of $V_0$ and $B_0$ as described in Ref.~\cite{Jollet2014}. In Fig.~\ref{delta-delta1_volume} it is shown how the $\Delta'$-factor is indeed very strongly correlated with the differences in equilibrium volume, $\delta V_0$, obtained from the equations of state (note that the data for $\Delta$-factor, $\Delta'$-factor and $\delta V_0$ reported in the scatter plots of Fig.~\ref{delta-delta1_volume} are taken between pairs of PSPs for the elemental crystals tested in this work and not between PSPs and all-electron results). Instead the $\Delta$-factor is scattered along a wide range of slopes that correspond to different values of the bulk modulus $B_0$: small (large) bulk moduli correspond to small (large) slopes.
In our protocol we adopt the $\Delta$-factor (and not the $\Delta'$-factor), as much more $\Delta$-factor reference data has been produced in the literature  \cite{Lejaeghere2016} for several PSP libraries and codes. However, for completeness, we compute also the renormalized $\Delta'$-factor and report this value as well. 
Broadly speaking two compared equations of state can be considered undistinguishable if the $\Delta$-factor is smaller than 1 meV/atom (valid for most of the elemental solids but with the notable exception of very soft materials) or if the $\Delta'$-factor is smaller than 3 meV/atom, where the latter corresponds to a variation in the equilibrium volume of less than 0.5\% for all elemental solids (see Fig.~\ref{delta-delta1_volume}).\\ 
The reference all-electron results of the equation of states chosen in this work are the ones of the WIEN2k code~\cite{Blaha1990} reported in Ref.~\cite{Lejaeghere2016}, with the exception of the rare-earth nitrides for which we use the WIEN2k results reported in Ref.~\cite{Topsakal2014}. All PSP calculations needed for the $\Delta$-factor estimation are performed at the reference wavefunction cutoff of 200 Ry using a dense Monkhorst-Pack~\cite{MonkhorstPack} k-grid of $20 \times 20 \times 20$ and a Marzari-Vanderbilt smearing~\cite{Marzari1999} of 2 mRy. Magnetism is included for the equations of state of oxygen and chromium (antiferromagnetism), manganese (antiferrimagnetism) and iron, cobalt, nickel and the rare-earth nitrides (ferromagnetism).

\begin{figure*}[!hbtp]
\centering
\includegraphics[scale=0.20]{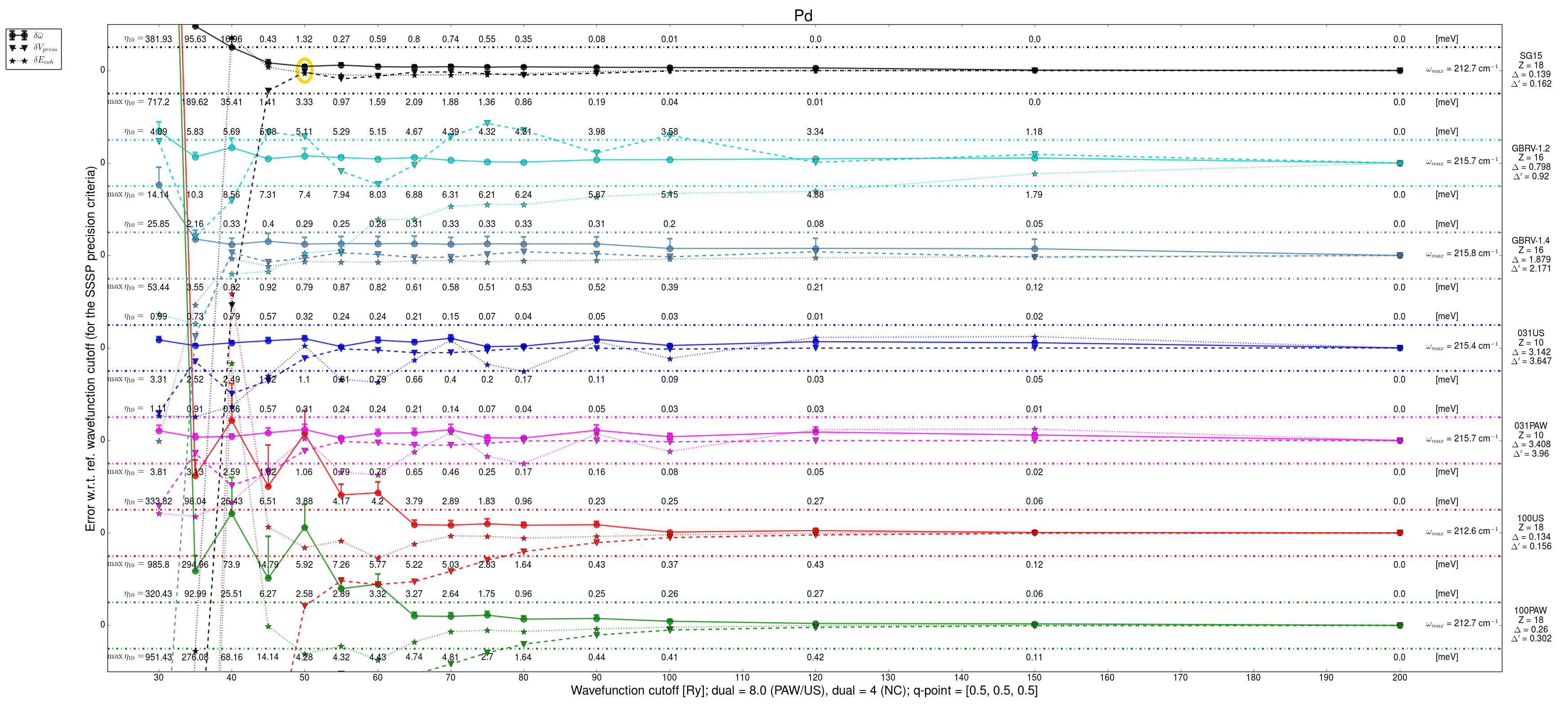}
\caption{SSSP testing protocol applied to palladium. For each pseudopotential the convergence w.r.t. the wavefunction cutoff of the zone-boundary phonons ($\delta \bar{\omega}$), cohesive energy ($\delta E_{coh}$), pressure ($\delta V_{press}$) and bands structure ($\eta_{10}$ and $\max \eta_{10}$) is monitored. The horizontal dashed lines correspond to the thresholds of the SSSP selection criteria (efficiency or precision); here precision is shown. On the right-hand side we report the number of valence electrons of the pseudopotential (Z), the $\Delta$-factor and the $\Delta'$-factor with respect to the reference all-electron results and the converged value of the highest phonon frequency ($\omega_{max}$). The circle marks the pseudopotential and wavefunction cutoff chosen for the SSSP library (version 1.1).\\
All convergence pattern plots of the 85 elements tested are available on the Materials Cloud platform~\cite{materialscloud_webpage}.}
\label{convergence-pattern_pd}
\end{figure*}

\noindent
Within the SSSP testing protocol we study the convergence of four different  quantities as a function of the wavefunction cutoff $E_c$, i.e. of the number of plane-waves used in the expansion of the Kohn-Sham states. The tested quantities are phonons frequencies at the zone-border, cohesive energies, pressure, and band structures.\\
All the calculations are performed on the ground-state structures of elemental crystals at 0 K, as provided in Ref.~\cite{cottenier_webpage}, with the exception of fluorine for which the SiF$_4$ structure is used because of convergence issues of the elemental fluorine structure and of lanthanides that are not included in the test set of Ref.~\cite{cottenier_webpage} and for which the nitride structures of Ref.~\cite{Topsakal2014} are used. In total we test 85 different elements of the periodic table.\\
In all PSP frameworks, a plane-wave representation of the charge density requires a cutoff, $E_{\rho}$, higher than the wavefunction cutoff, $E_c$~\cite{Note3}. 
Typically, convergence tests are performed by varying $E_c$ and keeping the dual, i.e. the ratio $E_{\rho} / E_{c}$, fixed. For instance, in the NC scheme the charge density is simply the modulus squared of the single-particle wavefunctions, summed over all the electrons, and in reciprocal space it reads:
\begin{equation}
\label{eq-rho}
\rho({\mathbf{G}}) = \sum_{n,\mathbf{k}}\sum_{\mathbf{G}'} \psi_{n,\mathbf{k}}({\mathbf{G}-\mathbf{G'}}) \psi^*_{n,\mathbf{k}}(\mathbf{G}')
\end{equation}
where sums run over the occupied bands with index $n$, Bloch vectors $\mathbf{k}$ and reciprocal lattice vectors $\mathbf{G}'$. Hence, the largest $\mathbf{G}$-vector appearing in the charge density has modulus twice  as  large  as  the  largest $\mathbf{G}$-vector appearing  in the wavefunction and, as plane-wave energies scale quadratically, the dual should be equal to 4 to guarantee that all Fourier components are represented. So for NC PSPs we always adopt a dual of 4, although it is known that in some cases calculations could be efficiently converged with a lower dual. 
In the PAW and US formalisms, wavefunctions are designed to be slowly varying in real space (i.e. to be soft) requiring substantially fewer plane-waves to be represented with respect to NC pseudo-wavefunctions. The price for working with a reduced basis set is the additional complexity in deriving expressions for observables, including the charge density which cannot be simply computed using Eq. \ref{eq-rho}. However the charge density $\rho$ is the fundamental quantity in DFT. The physics of the problem and the energy functional determine the spatial variation of $\rho$ and thus the cutoff $E_{\rho}$, independently of the PSP scheme that is adopted. Fundamentally, that is why PAW and US PSPs require relatively high duals, where according to common knowledge a choice of a dual of 8 is usually reasonable to efficiently achieve good precision. However, the convergence patterns at duals equal to 12 and 16 are also checked for selected elements that show a particularly high wavefunction cutoff, namely manganese, iron, cobalt, hafnium and oxygen, as also suggested by a convergence study on Fe~\cite{Dragoni2015}. \\
The convergence patterns are obtained at fixed duals as specified above and $E_c^{\text{ref}}=200$ Ry as the reference wavefunction cutoff~\cite{Note4}.
All the quantities are considered as differences with respect to the corresponding reference value calculated at $E_c^{\text{ref}}$. An example of the calculated convergence pattern plot is shown in Fig. \ref{convergence-pattern_pd} for the case of palladium.\\
We choose 200 Ry as the reference wavefunction cutoff because, for each element, all the quantities tested in the SSSP testing protocol typically convergence well before that value for at least one PSP (with radon being the only exception, as discussed in the Supplementary Figure 1). Therefore, even if for some hard PSPs the convergence plots could be marginally different by using a larger value for $E_c^{\text{ref}}$, those modifications would not be relevant for the conclusions of our work and, in particular, the selection of the SSSP libraries would not be affected.\\
We perform all the tests on the elemental crystals using a relatively coarse $6 \times 6 \times 6$ Monkhorst-Pack k-grid (except for oxygen and all the lanthanides where a $10 \times 10 \times 10$ k-grid is used instead) because in our protocol for convergence we are not directly interested in the absolute values of the tested quantities but rather on their difference with respect to the reference values computed at $E_c^{\text{ref}}$. We also disregard spin-polarization in all the convergence tests but we have verified for the magnetic structures that the convergence patterns are not substantially altered by the inclusion of magnetism (see Supplementary Figures 2-7).

\paragraph{Phonon frequencies.}
The convergence of vibrational properties of elemental crystals is performed by calculating, within the framework of DFPT, the phonon frequencies at the zone-border of the Brillouin zone, i.e. at the point $\bm{Q}=(\frac{1}{2},\frac{1}{2},\frac{1}{2})$ in relative coordinates of the reciprocal lattice vectors. While the $\Delta$-factor test is related to the structural and elastic properties of the system, by considering phonon frequencies at the border of the Brillouin zone we have access to information related to both acoustic and optical modes.\\
The number of phonon frequencies depends on the number of atoms in the unit cell, and so on the element under investigation. In the convergence pattern plots of the SSSP testing protocol we condense the information related to the several phonon frequencies into a single number $\delta \bar{\omega}$. It is defined as a relative average deviation (in percentage) among all the phonon frequencies $\omega_i$ calculated at $\bm{Q}$ for each wavefunction cutoff $E_c$:
\begin{equation}
\delta \bar{\omega} = \sqrt{\frac{1}{N}\sum_{i=1}^N \left| \frac{\omega_i(E_c) - \omega_i(E_c^{\text{ref}})}{\omega_i(E_c^{\text{ref}})} \right|^2}
\end{equation}

\noindent
where $N$ is the total number of phonon frequencies. The maximum relative deviation is similarly defined as
\begin{equation}
\delta \bar{\omega}_{error} = \max_i \left| \frac{\omega_i(E_c) - \omega_i(E_c^{\text{ref}})}{\omega_i(E_c^{\text{ref}})} \right|
\end{equation}

\noindent
and it is represented as an half error bar in the convergence pattern plots.\\
If the highest phonon frequency $\omega_{max}$ of an elemental crystal at $\bm{Q}$ is smaller than 100 cm$^{-1}$ at $E_c^{\text{ref}}$, the absolute average deviation and the corresponding maximum deviation are computed instead of the relative ones, since a precision of a few cm$^{-1}$ is often the reasonable target for a DFPT calculation.

\paragraph{Cohesive energies.}
We investigate the convergence of the energy difference between the crystalline solid and the corresponding individually isolated atoms, i.e. the cohesive energy of the elemental crystals. Since periodic boundary conditions are used in the calculations, the isolated atom is placed in a cell of lattice parameter equal to 12 $\text{\AA}$ to avoid spurious interactions with the periodic images. The quantity $\delta E_{coh}$ considered in the SSSP testing protocol is defined as the absolute difference between the cohesive energy at a given cutoff $E_c$ and the one at the reference wavefunction cutoff $E_c^{\text{ref}}$, i.e. 200 Ry (in units of meV per atom).

\paragraph{Pressure.}
We evaluate the convergence of the stress by computing the hydrostatic pressure, which is defined as $P=1/3 \, \text{Tr}(\sigma)$, where $\sigma$ is the stress tensor. 
Rather than checking convergence directly on the pressure itself (the magnitude of which depends strongly on the stiffness of the material) we evaluate it through its conversion into an equivalent volume. This allows the definition of a stiffness-agnostic and hence material's independent convergence criterion.
Starting from the Birch-Murnaghan equation of state for the pressure fitted on the reference all-electron calculations
\begin{equation}
\begin{split}
P_{\text{BM}}(V)= & \frac{3B_0}{2}
 \left[\left(\frac{V_0}{V}\right)^\frac{7}{3} - 
\left(\frac{V_0}{V}\right)^\frac{5}{3}\right]
\times \\ & \times  \left\{1+ \frac{3}{4}\left(B_0^\prime-4\right)
\left[\left(\frac{V_0}{V}\right)^\frac{2}{3} - 1\right]\right\},
\end{split}
\end{equation}
we define the deviation volume $V'$ as the one closest to the equilibrium volume $V_0$ such that $P_{\text{BM}}(V')=\delta P$ where $ \delta P = P(E_c) - P(E_c^{\text{ref}})$ is the residual pressure of a calculation performed at the cutoff $E_c$. With this definition, fully converged values of pressure give $\delta P = 0$ and therefore $V' = V_0$. Once $V'$ is known we can eventually find the relative volume deviation (in percentage) due to the residual pressure: $\delta V_{press}=(V'-V_0)/V_0$, which is the quantity considered in the SSSP testing protocol.

\paragraph{Band structure.}
The tests discussed so far deal with ground-state quantities only, computed either using DFT or DFPT. However, PSP calculations are often employed to study optical, transport and other properties that involve charged or neutral excitations. The majority of excited-states calculations are based on many-body perturbation theory (MBPT), e.g. G$_0$W$_0$ and self-consistent GW~\cite{Hybertsen1986,Onida2002,Reining2017}, the Bethe-Salpeter equation (BSE)~\cite{Strinati1988}, or dynamical mean field theory (DMFT)~\cite{Georges1996}, and are performed on-top of a DFT calculation, which provides the starting point for both self-consistent and one-shot approaches. Hence, we include band structures in our testing protocol, taking into account both the occupied bands and some of the lower lying unoccupied bands. Here, we outline a protocol for performing both convergence tests and verification of band structures by defining a bands distance (a similar idea has been proposed independently in Ref.~\cite{Huhn2017}). The aim is to quantify how much two band structures ``differ'' by introducing a simple and computationally inexpensive metric in the band structures space.
We call our bands distance $\eta$ and consider two cases that are distinguished solely by the number of bands taken into account. The $\eta_v$ (or ``eta valence'') considers the occupied bands only, while in the $\eta_{10}$ (or ``eta conduction 10'') all the bands up to $10$ eV above the Fermi level are considered.
We always use a robust (0.3 eV) Fermi-Dirac smearing to deal with partially occupied bands, while to compute $\eta_v$ for insulators we use no smearing. We choose a $6 \times 6 \times 6$ uniform k-grid, in the full Brillouin zone and with no symmetry reduction. Choosing a high-symmetry path could result in an unsatisfactory arbitrary choice, as different recipes for the standardization of paths have been introduced in the recent literature~\cite{pizzi2017,curtarolo2010} and interesting features of the band structure may occur far from the high-symmetry lines (such as Weyl points)~\cite{Xu2015,Soluyanov2015}. A uniform mesh is also more appropriate from the point of view of electron's nearsightedness~\cite{Kohn2005}: if the energy eigenvalues are known on a sufficiently fine uniform $\mathbf{k}$-points mesh, it is possible to get an \emph{exact} real-space representation of the Hamiltonian in a Wannier function basis \cite{marzari2012} and then interpolate to an arbitrary fine mesh.\\
Let us suppose we have two sets of bands $\epsilon^A_{n\mathbf{k}}$ and $\epsilon^B_{n\mathbf{k}}$; we define the distance between the two sets of (valence) bands as
\begin{equation}
\label{eta_def}
\eta_{v} (A,B)= \min_{\omega} \sqrt{ \frac{ \sum_{n\mathbf{k}}\tilde{f}_{n\mathbf{k}}(\epsilon^A_{n\mathbf{k}} -  \epsilon^B_{n\mathbf{k}} + \omega )^2  }{\sum_{n\mathbf{k}} \tilde{f}_{n\mathbf{k}}   } },
\end{equation}
where
\begin{equation}
\tilde{f}_{n\mathbf{k}}=\sqrt{f_{n\mathbf{k}}(\epsilon^A_{ {F}},\sigma) f_{n\mathbf{k}}(\epsilon^B_{ {F}},\sigma) },
\end{equation}
\\
$ f_{n\mathbf{k}}(\epsilon,\sigma)$ being the Fermi-Dirac distribution and $\sigma$ the smearing width. The Fermi energies, $\epsilon^{(A,B)}_{ {F}}$, for the two band structures $A$ and $B$ are obtained from the relation $N^{(A,B)}_{el} = \sum_{n\mathbf{k}} f^{(A,B)}_{n\mathbf{k}}(\epsilon^{(A,B)}_{ {F}},\sigma)$, where $N_{el}$ is the number of electrons.  In order to properly align the two sets of bands,  $\eta_{ {v}}$ is defined as the minimum with respect to a rigid energy shift $\omega$.\\
We now consider also the low lying conduction bands by introducing $\eta_{10}$, defined as in Eq. \ref{eta_def} but with a Fermi level up shift of 10 eV. In this way,  $\eta_{10}$  measures the bands distance of the valence bands plus the conduction bands up to $10$ eV above the Fermi energy.\\
Finally, we also take into account the possibility that significant differences between band structures may occur only in subregions of the Brillouin zone or in small energy ranges. After computing the $\eta$, we check the slowest converging band by computing max $\eta$, defined as
\begin{equation}
\text{max} \,\, \eta = \max_{n\mathbf{k}} |\epsilon^A_{n\mathbf{k}} -  \epsilon^B_{n\mathbf{k}} + \omega |,
\end{equation}
and request that is has to be converged with a slightly higher threshold than $\eta$ itself.\\
In the SSSP testing protocol we use $\eta_{10}$ and max $\eta_{10}$ (in units of meV) as criteria to quantitatively study the convergence of band structures.\\

\section{Discussion}

\subsection{SSSP selection criteria}
\noindent
We discuss now the selection criteria used to build our optimal PSP libraries, namely the SSSP efficiency and SSSP precision libraries (version 1.1).
As mentioned in the introduction, our primary goal is to define tested PSP libraries with a focus on efficiency and precision for high-throughput calculations and to suggest converged wavefunction cutoffs. 
The main idea behind the SSSP precision library is to provide the PSPs that are the closest to all-electron calculations in terms of $\Delta$-factor computed at the reference wavefunction cutoff $E_c^{\text{ref}}$, without much consideration on the computational cost and the wavefunction cutoffs actually needed to converge all relevant quantities. On the other hand, the SSSP efficiency library is designed for practical applications that should remain affordable, and therefore PSPs are chosen such that wavefunction cutoffs are as low as possible while keeping the precision reasonable.\\
The selection criteria are listed in Table~\ref{criteria:tab}. 
For SSSP efficiency, when possible we select PSPs with a rather small $\Delta$-factor (below $1$~meV/atom). The phonons $\delta \bar{\omega}$ should be converged within $2 \%$ (or within 2 cm$^{-1}$ if the highest phonon frequency is smaller than 100 cm$^{-1}$), the cohesive energy $\delta E_{coh}$ within $2$~meV/atom, the pressure within $1 \%$ for $\delta V_{press}$ (i.e. 0.33\% on the lattice parameter of a cubic crystal) and the band structure within 10 meV for $\eta_{10}$ and within 20 meV for $\max \eta_{10}$. For the SSSP precision, the criteria are slightly stricter (see Table~\ref{criteria:tab}) and we systematically opt for the PSP with the smallest $\Delta$-factor. Therefore the wavefunction cutoffs of the SSSP precision are typically higher than the ones proposed for the SSSP efficiency.\\
We underline here that in a few difficult cases the SSSP libraries are built following these criteria as general guidelines and not using these as strict rules. In practice, this means that the PSPs are chosen one-by-one through human inspection and not with an automatic procedure. This flexible approach is necessary because the convergence of some of the tested quantities is sometimes slow and/or irregular. For example it can happen that the selection criteria are not all together satisfied at a reasonable wavefunction cutoff for any of the PSPs of a given element or that the convergence patterns show outlier data points or oscillations.
A clear example of this situation is given by the extremely soft noble-gas elemental solids for which the convergence of the tested quantities, in particular of the stress tensor and of the phonon frequencies, can be very noisy due to numerical instabilities.
In these cases it is therefore necessary to make compromise choices that can sacrifice or increase some of the thresholds imposed by the SSSP selection criteria, if no other possibilities are available or in order to keep the computational cost reasonable.\\

\subsection{Ghost states}
\begin{figure*}[!hbtp]
\centering
\includegraphics[scale=1.15]{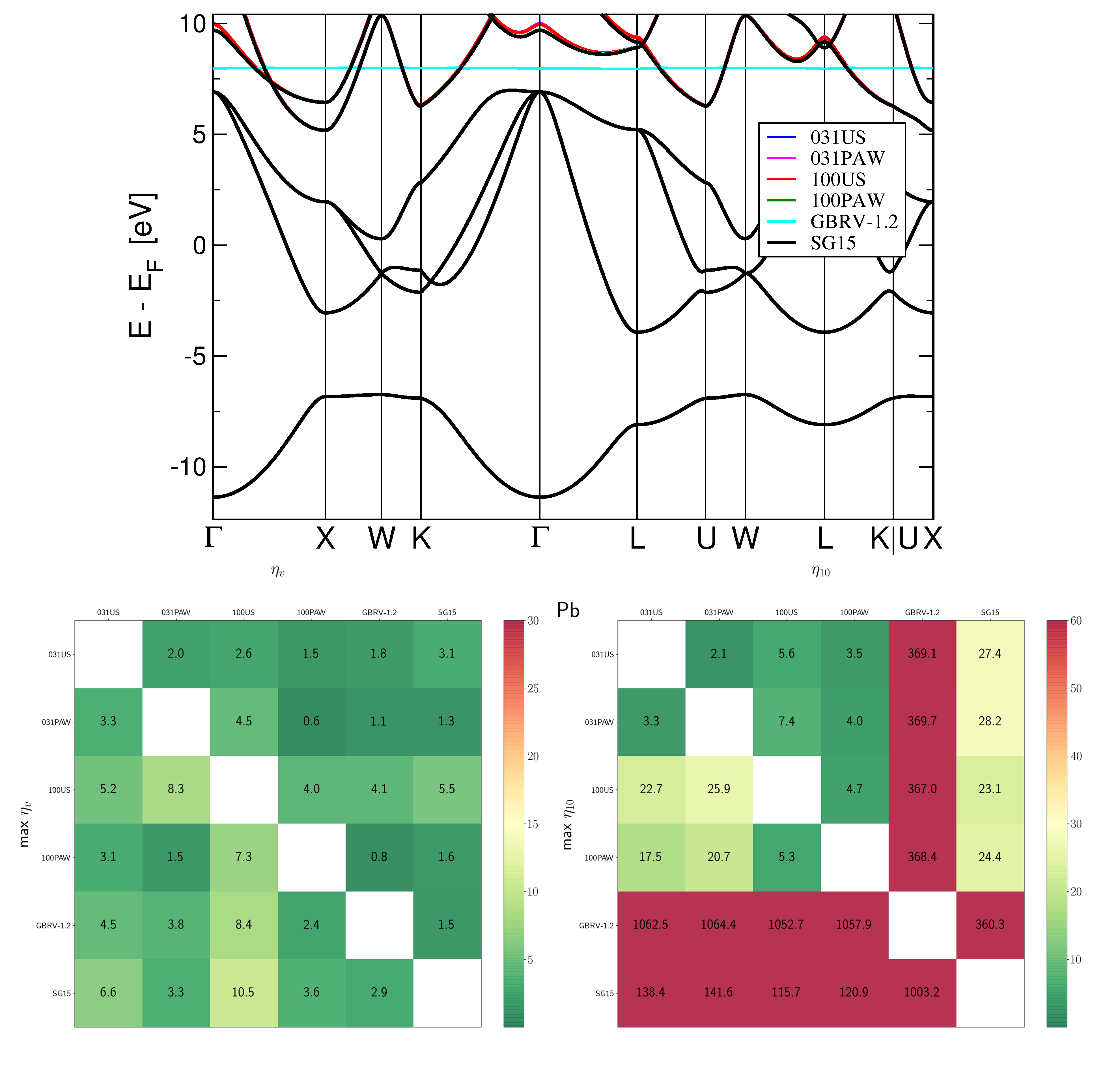}
\caption{\label{chess}Band structure of FCC Pb along a high-symmetry path, for several pseudopotential libraries (top panel). The valence bands are almost identical to each other, while some differences appear in the conduction bands: the SG15 bands deviate from the other bands around 7-10 eV over the Fermi level and a flat ghost state in the GBRV bands is clearly visible at around 8 eV. These differences between band structures can be compressed into the bands distances $\eta_v$ and $\eta_{10}$, reported in units of meV (bottom panel).  In addition, ghost states in the band interval considered can be automatically detected as peaks in the $\eta$ function, hence simplifying greatly the verification of spectral properties.}
\end{figure*}

\noindent
We use the bands distance $\eta_{10}$ defined above not only for the convergence tests but also to compare the band structures of the tested PSPs for all the elemental crystals considered. However different PSPs are often generated with different combinations of semi-core states in the valence band. Hence, we compare only the bands they have in common, by taking the minimum number of electrons of all the set and cutting the exceeding low-energy bands accordingly~\cite{Note5}. \\
By means of this additional criterion it is possible to automatically detect ghost states~\cite{Gonze1991} in a PSP in the valence and in the conduction up to the chosen threshold (here 10 eV above the Fermi energy), as they are signalled by extremely large values (of the order of eV or more) of the bands distances when computed with respect to other ghost-free PSPs (see Fig.~\ref{chess} for an example).
A list of the tested PSPs having ghost states in the empty conduction bands is reported in Table~\ref{ghost_states}. However, it is worth noting that standard DFT calculations for ground-state properties are unaffected by ghost states above the Fermi level. Nonetheless they could be a possible source of problems for applications related to excited-state properties (e.g. in MBPT calculations such as GW or BSE). As expected, none of the PSPs considered has ghost states in the valence as they would give unphysical results also for ground-state properties and they would be easily spotted.
\\
We stress here that the bands distance could in principle also be used for verification studies because it would allow for a quantitative comparison of PSPs band structures with reference all-electron band structures.

\subsection{Correlations among tested quantities}
\begin{figure*}[!hbtp]
\centering
\includegraphics[scale=0.65]{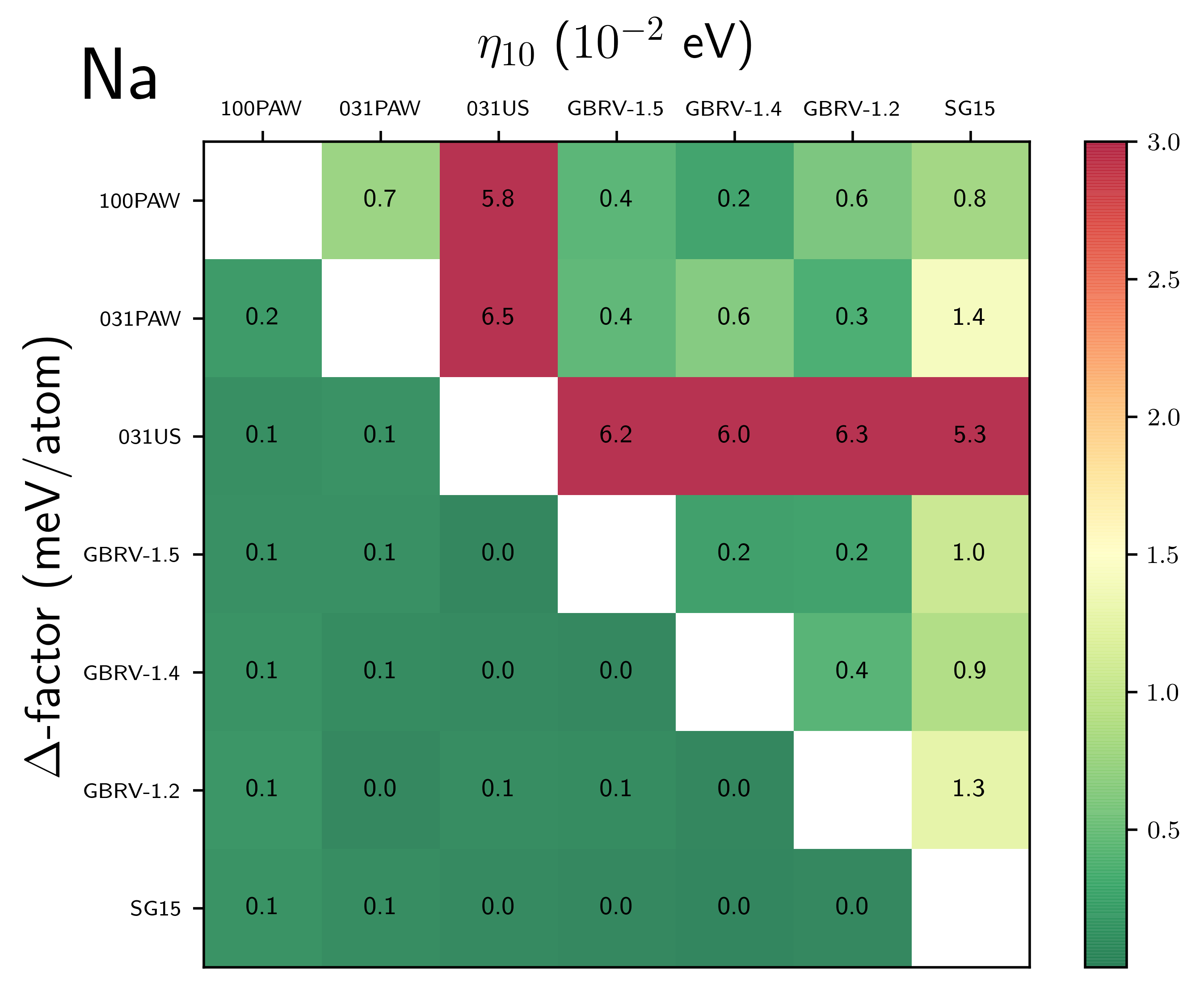}
\caption{Comparison of the discrepancies among PSPs of sodium for the equations of state ($\Delta$-factor) and band structures ($\eta_{10}$) at the reference wavefunction cutoff of 200 Ry. Although all equations of state are very similar among each other, the 031US band structure shows some discrepancies with respect to other PSPs.}
\label{sodium_eta10-delta_chess}
\end{figure*}

\begin{figure*}[!hbtp]
\centering
\includegraphics[scale=0.8]{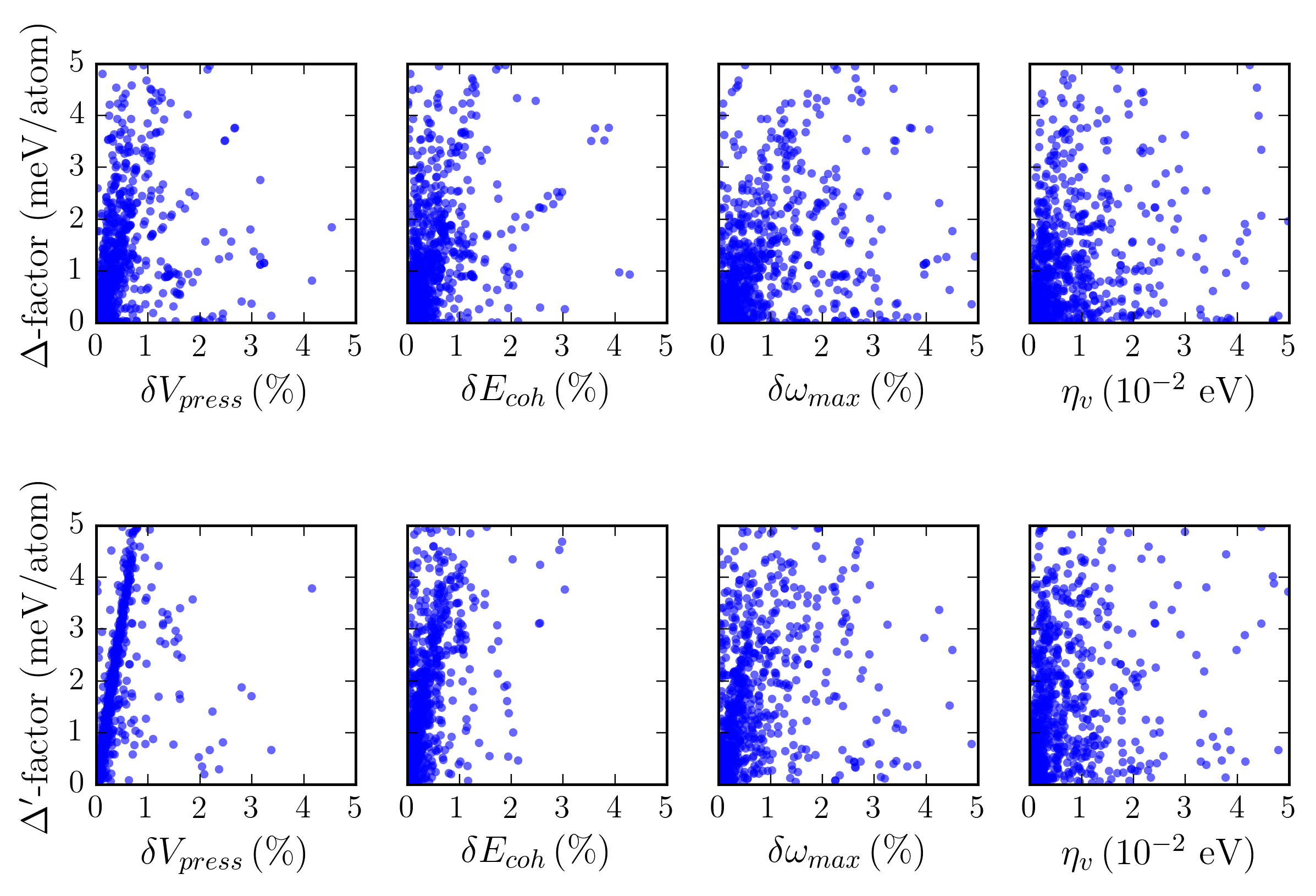}
\caption{Scatter plots showing the discrepancies among PSPs for all the 85 elements tested. Each point in the plot corresponds to the discrepancies between pairs of PSPs (of a given element) for the two quantities tested calculated at $E_c^{\text{ref}}=200 \text{ Ry}$. In particular we show the comparisons between the equations of state ($\Delta$-factor in the upper panel, $\Delta'$-factor in the lower panel) and, respectively, the pressure ($\delta V_{press}$), cohesive energy ($\delta E_{coh}$), highest phonon frequency ($\delta \omega_{max}$) and valence band structure ($\eta_v$). No strong correlations among the quantities tested are observed but we note that the $\Delta'$-factor is more correlated to $\delta V_{press}$ than the $\Delta$-factor.}
\label{scatter-plots}
\end{figure*}

\begin{figure*}[!hbtp]
\centering
\includegraphics[scale=0.65]{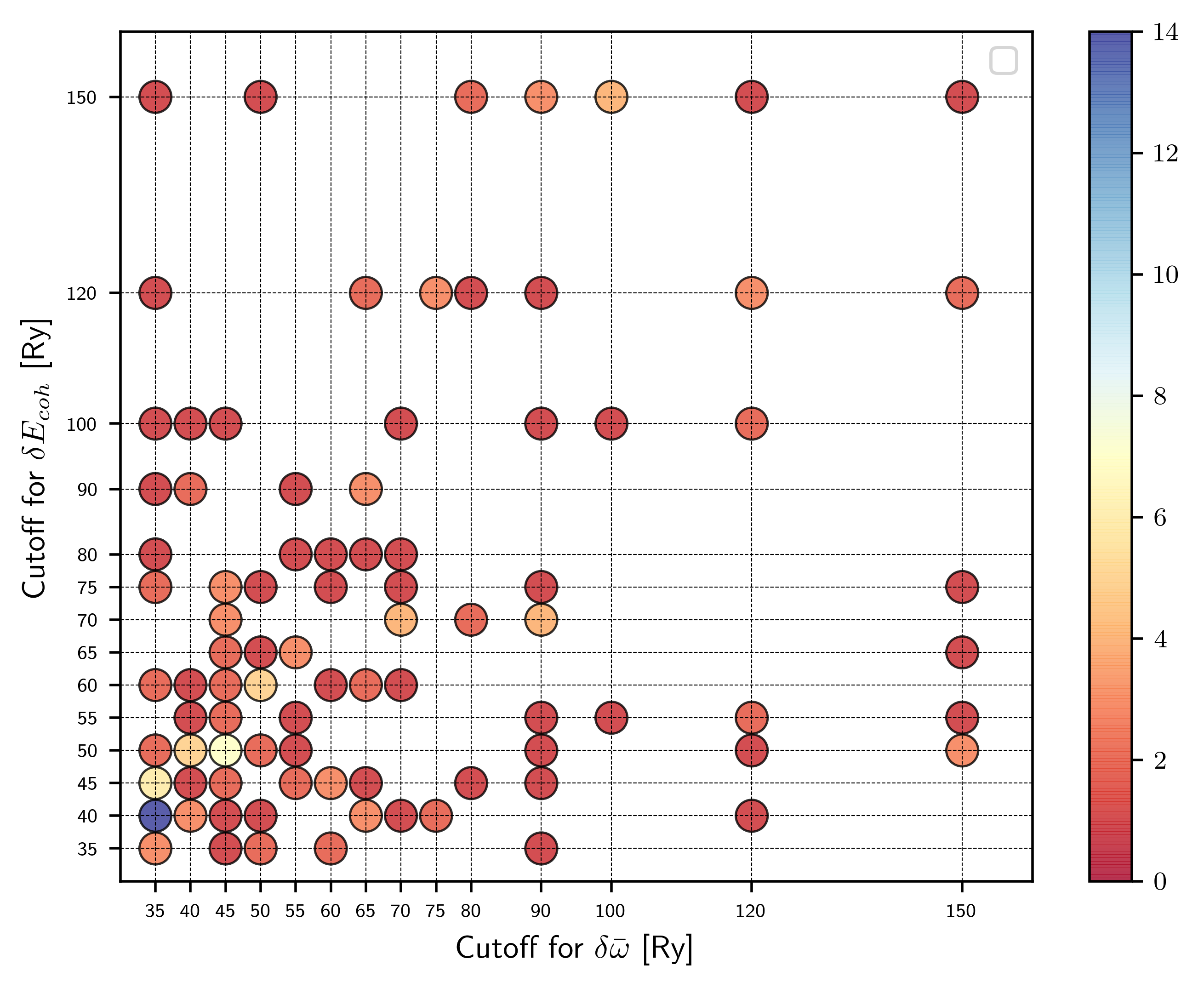}
\caption{Comparison of the  PSP wavefunction cutoffs (circles) selected strictly following the SSSP efficiency selection criteria (see Table~\ref{criteria:tab}) for the phonon frequencies ($\delta \bar{\omega}$) and the cohesive energies ($\delta E_{coh}$).  The colormap for the circles corresponds to the number of PSPs that have that pair of wavefunction cutoffs for the two quantities tested. We find that, in general, the convergence of one quantity does not imply convergence of the other.}
\label{cutoffs-scatter}
\end{figure*}

\noindent
Before giving the list of PSPs chosen by following the SSSP testing protocol, we show with an analysis of our results that an accurate selection of PSPs for generic applications in computational material science needs several independent criteria to be satisfied, based on the estimation of different physical properties. This is done also with the purpose of furnishing an \textit{a posteriori} justification of the protocol we established for PSP testing. In particular we show that PSPs that give very similar results for a certain tested quantity can give, in a non-negligible number of cases, significant discrepancies in the estimation of some other quantity. \\
We compare the PSP results for the physical properties considered in the SSSP testing protocol by calculating the discrepancies between all the available PSPs for a given element, using the data obtained at the reference wavefunction cutoff of 200 Ry. An example is shown in Fig.~\ref{sodium_eta10-delta_chess} where the differences between the equations of state (through the use of the $\Delta$-factor) and between band structures (through the use of $\eta_{10}$) are compared for the case of sodium. We notice that, even if all the sodium PSPs we considered produce very similar equations of states---with $\Delta$-factors always smaller than $0.3$ meV---the band structure of a particular PSP shows instead substantial differences up to $\eta_{10}=65$ meV.\\ 
More generally, we can study correlation between pairs of quantities by looking at scatter plots where the differences between all possible couples of PSPs for all the 85 elements are considered. 
In Fig.~\ref{scatter-plots} we show as an example the correlation plot between the equation of state (obtained through both the $\Delta$-factor and the $\Delta'$-factor) and pressure ($\delta V_{press}$), cohesive energy ($\delta E_{coh}$), highest phonon frequency ($\delta \omega_{max}$) and valence band structure ($\eta_v$). 
Fig. \ref{scatter-plots} shows how for all such properties correlation is very weak, suggesting that the precision of a PSP is property dependent.
However, we also notice (see Fig.~\ref{scatter-plots}) that the correlation between the $\Delta'$-factor and $\delta V_{press}$ is higher than for the $\Delta$-factor. This observation can be rationalized in terms of the $\Delta'$-factor renormalization, that provides an estimate of the difference between two equations of state that is more material-independent and straightforwardly related to volume differences (see Fig.~\ref{delta-delta1_volume} and discussion therein). 
From our results we conclude that there is no strong correlation between pairs of tested quantities entering our selection criteria, hence the similarity between PSPs is strongly property dependent.\\
Up to now, the $\Delta$-factor is the only verification test present in the literature to assess the precision of DFT calculations of solids. Given the small correlations among the SSSP criteria, we stress here the importance for the electronic structure community to head for the creation of an heterogeneous set of validated all-electron reference data, which would ideally include other physical properties beyond the equation of state, such as phonons or band structures (such effort is currently coordinated by S. Cottenier). This would allow the extension of the available PSP verification tests beyond the $\Delta$-factor for elemental crystals, potentially improving the assessment of PSPs precision. \\
Similarly, we show that the convergence with respect to the wavefunction cutoff of a given physical property usually has a different and uncorrelated behaviour if compared to other tested quantities (see Fig.~\ref{cutoffs-scatter} for a comparison of the wavefunction cutoffs at which two tested quantities in the SSSP testing protocol reach the required precision for each PSP). Indeed, differences in the mathematical expression adopted and/or in the code-specific implementations that are needed to compute the tested quantities can result in different and independent convergence patterns, so that each quantity reaches the required precision at different wavefunction cutoffs. For example, the derivatives involved in the calculation of the stress tensor or the phonon frequencies either through direct, finite differences methods or linear response theory, can introduce different numerical noise and display a slower convergence if compared to other properties, such as equations of state or band structures, that do not require calculations of derivatives.\\
In general, it is therefore necessary to study the convergence of each relevant quantity separately, in order to correctly estimate the optimal number of plane waves that gives results converged within the required precision for all the properties of interest.\\


\subsection{Exchange-correlation functionals}

\begin{figure*}[!hbtp]
\centering
\includegraphics[scale=0.19]{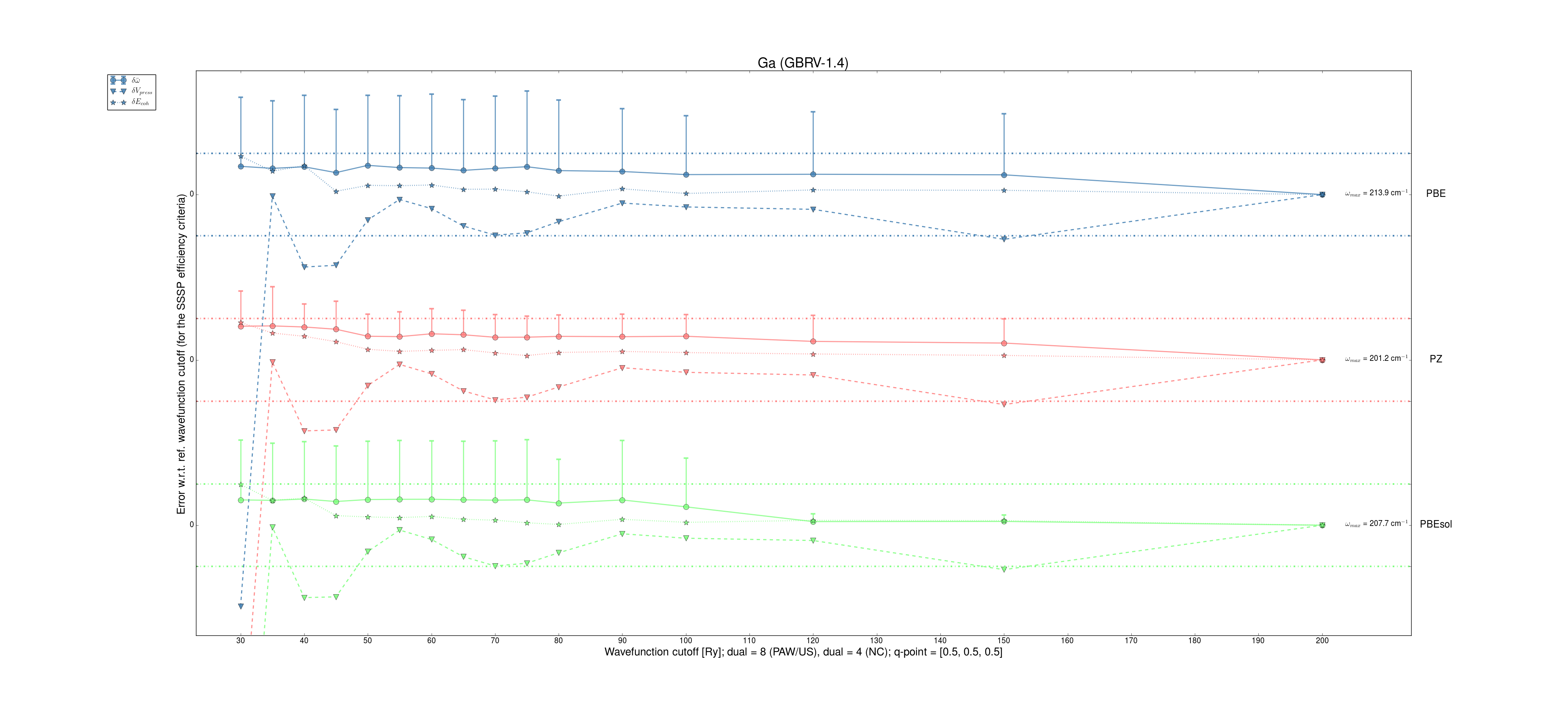}
\caption{Comparison of the convergence pattern plots for three functionals, i.e. PBE, PBEsol and PZ, applied to the GBRV-1.4 PSP of Ga. PBEsol and PZ PSPs are generated with the same atomic parameters of the original PBE pseudopotential from the GBRV-1.4 PSP library.}
\label{Ga_functional}
\end{figure*}

\noindent
Our study on the precision and efficiency of PSPs is restricted to the PBE functional as it is among the most popular ones in the electronic structure community and the only functional for which a verified set of reference all-electron results for solids exists~\cite{Lejaeghere2016}.
So the $\Delta$-factor test for verification can be performed, at the moment, only with the PBE functional. \\
Still, it is worth to comment on the transferability of the convergence tests performed in the SSSP testing protocol among different functionals.
For this purpose, we consider the revised PBE GGA for solids, namely the PBEsol~\cite{Perdew2008} functional, and one of the most widely used functionals for the local density approximation (LDA), i.e. the PZ~\cite{Perdew1981} functional.  
By testing some elemental crystals for the GBRV library (see Fig.~\ref{Ga_functional} for the case of Ga and the Supplementary Figures 8-9 for a few more systems), we find that the convergence patterns turn out to be very similar if the PSPs are generated with the same atomic parameters (such as electronic configuration, cutoff radii, etc.), thus showing a good transferability of the convergence tests among different local and semi-local functionals.
However, performing consistent tests for the transferability of more complex and nonlocal functionals, such as SCAN~\cite{Sun2015} (meta-GGA) or HSE~\cite{Heyd2003} (exact-exchange), is less straightforward because, as of today, no PSP libraries for these kind of functionals exist.
Indeed, in these cases, the common approach followed in all PSP DFT codes is to use PSPs generated with local or semi-local functionals, e.g. PBE, and to then ``switch" to the complex functional, e.g. HSE, when performing the PSP DFT calculation including the valence electrons only. Tests performed following this approach, although useful, would not ensure the transferability of the functional under investigation and more extensive and consistent studies on the subject are therefore left to future work, in particular when HSE or SCAN PSP libraries, to only cite a few possible examples, will be available.\\


\subsection{SSSP libraries}
\begin{figure*}[!hbtp]
\centering
\includegraphics[scale=0.4]{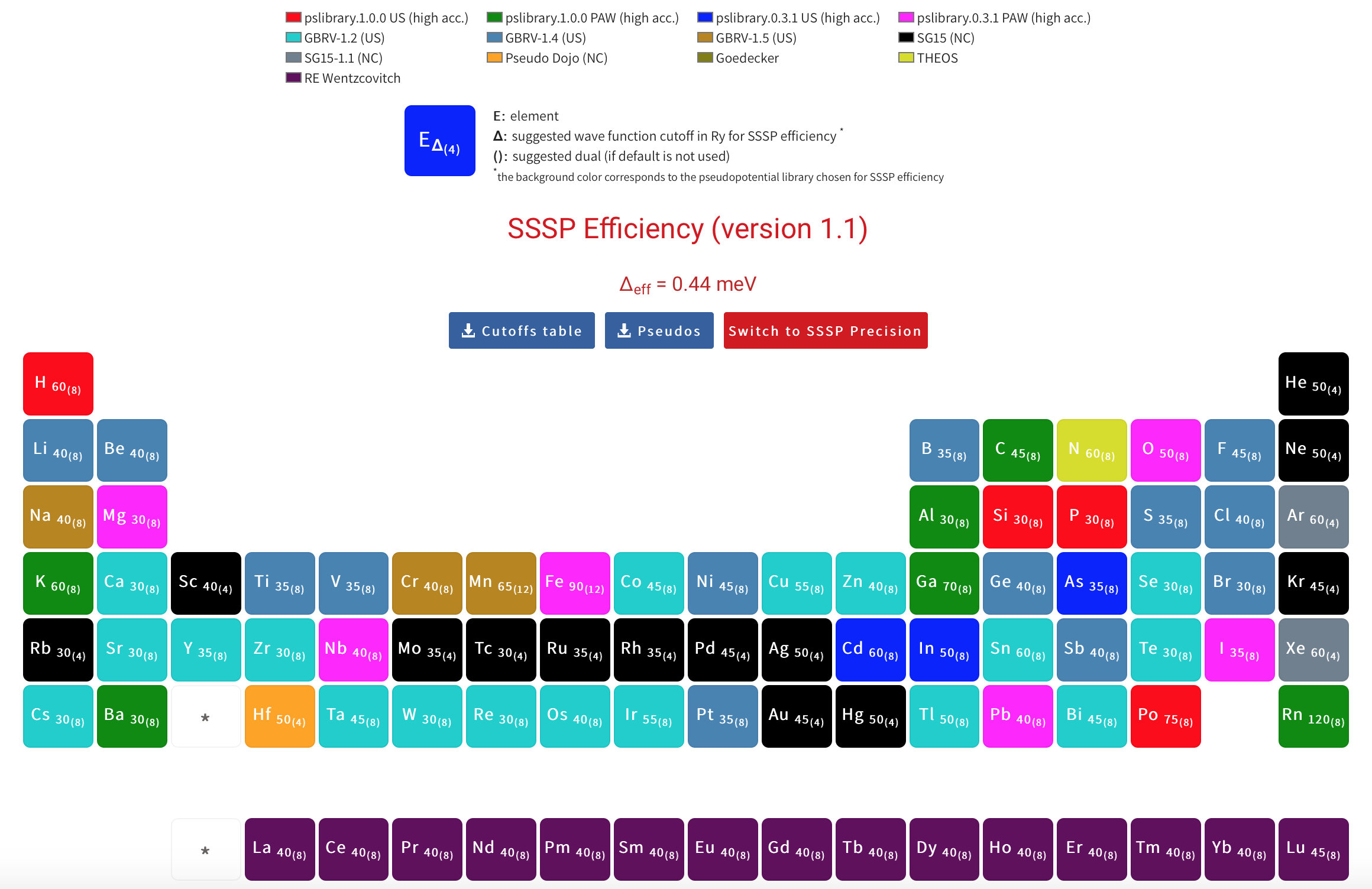}
\caption{SSSP periodic table (here efficiency is shown) from the Materials Cloud webpage~\cite{materialscloud_webpage} (version 1.1). By clicking on an element, the data and results of the SSSP testing protocol (convergence pattern plots as in Fig.~\ref{convergence-pattern_pd}, equations of state, band structures and more) for the selected element are accessible.}
\label{sssp_periodic-table}
\end{figure*}

\noindent
Table~\ref{sssp_pseudo} and Table~\ref{sssp_pseudo1} show the two SSSP libraries, efficiency and precision (version 1.1), selected according to the SSSP selection criteria specified above. The suggested wavefunction cutoffs (in Ry) and the duals are also indicated for each PSP chosen. The SSSP periodic table with all PSPs, wavefunction cutoffs and duals is also accesible interactively online on the Materials Cloud platform~\cite{materialscloud_webpage} (see Fig.~\ref{sssp_periodic-table}).  The average suggested wavefunction cutoffs of the SSSP efficiency and SSSP precision over all the 85 elements tested are 44 Ry and 56 Ry, respectively. A dual of 8 has been used for all PSPs except norm-conserving ones where a dual of 4 is used, and iron and manganese for which a dual of 12 is suggested.\\ 
The SSSP efficiency and SSSP precision have small average $\Delta$-factors of 0.44~meV and 0.33~meV, respectively (where the average is performed over all elements tested excluding the 15 rare-earth nitrides, following the recipe of Ref.~\cite{Lejaeghere2016}). \\ 
The SSSP libraries have already proven to be a reliable tool for a number of computational studies: for instance the beta version (called version 0.7) of the SSSP libraries have enabled the high-throughput computational exfoliation of two-dimensional materials~\cite{Mounet2018} and have supported the combined experimental and theoretical study of catalysts for oxygen evolution reaction~\cite{Lebedev2017}.\\
On a more general level, apart from the SSSP testing protocol and libraries, our work provides a database of verification data and convergence tests that facilitates the optimal choice of PSPs and wavefunction cutoffs for custom applications. For example, some physical properties may be implemented only for some PSP types (typically only NC) or some applications may require convergence of just a subset of the quantities that we consider in the SSSP testing protocol. By a look at our plots and data, see for instance the condensed plot for palladium shown in Fig.~\ref{convergence-pattern_pd}, a user can quickly select the optimal PSP and wavefunction cutoff tailored for the specific application. \\

\noindent
In summary, we propose an extensive testing protocol for PSPs to investigate precision and performance of several NC, US and PAW PSP libraries that are publicly available. We incorporate in the SSSP protocol a verification part, based on the $\Delta$-factor, and an efficiency part, based on the plane-wave cutoff convergence tests for phonon frequencies, cohesive energies, pressures, and band structures. Leveraging the SSSP protocol, we identify two optimal PSP libraries, named SSSP efficiency and SSSP precision (version 1.1), that provide thoroughly tested and precise PSPs for 85 elements of the periodic table, selected from publicly available PSP libraries~\cite{Kucukbenli2014,DalCorso2014,Topsakal2014,Garrity2014,Schlipf2015,Dojo2017,Willand2013}, for which the original authors should be acknowledged. Our effort not only is particularly relevant for high-throughput computational materials screening, where the right compromise between precision and computational cost is essential, but it substantially contributes to set high the bar of the quality of PSP calculations of solid-state materials. Building on the invaluable work behind all the PSP libraries we considered, we provide 
a systematic survey of PSP quality across multiple physical properties and multiple libraries and techniques (NC, US and PAW). 
Our work shows how the assessment of both precision and efficiency of pseudopotentials is strongly property-dependent and requires a multi-dimensional quality gauge, pointing to the need of a verification standard in the computational solid-state community. Given the importance of PSP calculations in modern materials science, nanotechnology, chemistry and physics, our findings call for more verification efforts aimed at increasing precision and efficiency of computed quantities that are routinely used to discuss novel physics, to help interpreting experiments or even to discover and design novel materials. In particular, we hope that this work will stimulate further investigations in the all-electron community, that ideally would provide more reference data for an heterogeneous set of properties elaborating on our discussion.

\section{Methods}
\noindent
All the calculations needed for this work (more than 50'000 DFT and DFPT calculations) were performed with the goal to ensure reproducibility of all the data obtained, compliant with the FAIR guiding principles for data management~\cite{Wilkinson2016}. This is the reason why we used AiiDA~\cite{Pizzi2016}, an open-source Python infrastructure for computational science, that is specifically designed to track the provenance of data and calculations and that allow the user to implement workflows that can run complex sequences of calculations. It is therefore particularly suited for high-throughput studies, such as the deployment of the SSSP testing protocol, where a large number of simulations are involved.\\ 
In practice AiiDA can prepare and submit calculations (usually to an HPC cluster) and then retrieve and store the results inside a database, all automatically. The database can be subsequently queried by the user to extract data or other useful informations.\\
The complete SSSP testing protocol is implemented as an AiiDA workflow, called SsspWorkflow, that can run all the convergence tests and the $\Delta$-factor verification test. The SsspWorkflow is built on top of the PwWorkflow, a very robust lower-level workflow in charge of handling all the QE simulations and that can restart calculations in case of standard QE errors or, for example, if the user-specified wall time is too small.\\ 
The SsspWorkflow allows a generic user to perform all the calculations required by the SSSP protocol in a completely automatic way. With this tool, other and new pseudopotential libraries could be easily tested in the future in order to update the subsequent versions of the SSSP libraries with more precise and efficient PSPs.\\




\section*{Acknowledgements}
\noindent
The authors warmly thank Fernando Gargiulo, Snehal Waychal and Elsa Passaro for assistance with the SSSP section of the Materials Cloud web platform, Giovanni Pizzi and Andrea Cepellotti for the help in developing AiiDA workflows, Marco Gibertini for numerous discussions and for pointing out several useful references, Nicolas G. H\"{o}rmann for further testing on the SSSP pseudopotential libraries, Stefaan Cottenier for critical reading of the manuscript.

\section*{Competing Interests}
\noindent
The authors declare no competing interests.

\section*{Author contributions}
\noindent
N. Ma. and I. C. designed the study; G.P., A.M., I.C. and N. Mo. developed the workflows and performed the calculations; G. P., A. M. and I. C. wrote the manuscript, and all authors discussed the protocol, analysed the data, and commented on the manuscript. G. P., A. M. and I. C. contributed equally to this work.

\section*{Funding}
\noindent
This work has been supported by NCCR MARVEL and by H2020 CoE MaX, computing time has been provided by the Swiss National Supercomputing Centre (CSCS) and by PRACE (Project Ids 2016153543 and 2016163963).
We also thank Sadas Shankar and Intel Corporation for early support to this project, through their seed funding on ``Validated
pseudopotentials for electronic-structure simulations" (2009-11).

\section*{Data availability}
\noindent
All the data produced in this work is freely available on the Materials Cloud online platform~\cite{materialscloud_webpage}, where the user can interactively browse the results and explore the data provenance. The full database with all the data can also be downloaded~\cite{sssp_archive}.

\section*{Additional information}

\noindent
\textbf{Supplementary information} is available at \textit{npj Computational Materials} website.\\


\section{References}
\bibliographystyle{naturemag}

\clearpage
\newpage

\section{Figures}

\clearpage
\newpage

\section{Tables}

\begin{table*}[!hbtp]
  \begin{center}
    \begin{tabular}{|c|c|c|c|}
      \hline
      \hline
      Full name & Short name & Method & Reference\\
      \hline
      pslibrary.0.3.1 US & 031US & US & \cite{Kucukbenli2014} \\
      pslibrary.0.3.1 PAW & 031PAW & PAW & \cite{Kucukbenli2014} \\
      pslibrary.1.0.0 US (high acc.) & 100US & US & \cite{DalCorso2014} \\
      pslibrary.1.0.0 PAW (high acc.) & 100PAW & PAW & \cite{DalCorso2014} \\
      GBRV-1.2 (US) & GBRV-1.2 & US & \cite{Garrity2014} \\
      GBRV-1.4 (US) & GBRV-1.4 & US & \cite{Garrity2014} \\
      GBRV-1.5 (US) & GBRV-1.5 & US & \cite{Garrity2014} \\
      SG15 (NC) & SG15 & NC & \cite{Schlipf2015} \\
      SG15-1.1 (NC) & SG15-1.1 & NC & \cite{Schlipf2015} \\
      RE Wentzcovitch (PAW) & Wentzcovitch & PAW & \cite{Topsakal2014} \\
      Goedecker (NC) & Goedecker & NC & \cite{Willand2013} \\
      PseudoDojo (NC) & Dojo & NC & \cite{Dojo2017} \\
      THEOS (US) & THEOS & US &  \\
      \hline
    \end{tabular}
    \caption{Pseudopotential libraries tested with the SSSP protocol. The short names correspond to the name used in the convergence pattern plots (see Fig.~\ref{convergence-pattern_pd}) to identify the pseudopotential libraries.}\label{pseudo_libraries}
  \end{center}
\end{table*}

\begin{table*}[!hbtp]
  \begin{center}
    \begin{tabular}{|l|c|c|c|}
      \hline
      \hline
      & SSSP efficiency & SSSP precision & notes \\
      \hline
      Phonon frequencies  ($\delta \bar{\omega}$) & $<2\%$ & $ <1 \% $ & $<1\text{ cm}^{-1}\text{ if }\omega_{max} <100 \text{ cm}^{-1} $  \\
      Cohesive energy ($\delta E_{coh}$) & $< 2$~meV/atom & $< 2$~meV/atom &  \\
      Pressure ($\delta V_{press}$) & $<1\%$ & $<0.5\%$ & in terms of volume differences \\
       Band structure ($\eta_{10}$) & $< 10$~meV & $< 10$~meV & \\
       Band structure ($\max \eta_{10}$) & $< 20 $~meV & $< 20$~meV & \\
       Equation of state ($\Delta$-factor) & $< 1$~meV/atom (if possible) & smallest &  \\
       \hline
    \end{tabular}
    \caption{Selection criteria for the SSSP efficiency and SSSP precision libraries.}\label{criteria:tab}
  \end{center}
\end{table*}

{ \color{red}
\begin{table*}[!hbtp]
  \begin{center}
    \begin{tabular}{|c|c|c|c|}
      \hline
      \hline
      Element & Pseudopotential \\
      \hline
      Ar & SG15 \\
      Cs & SG15 \\
      In & SG15 \\
      Hg & 031PAW \\
      Hg & 031US \\
      S & SG15 \\
      Se & SG15 \\
      Sn & SG15 \\
      Te & SG15 \\
      Pb & GBRV-1.2 \\
      Po & 031PAW \\
      Po & 031US  \\
      Sb & SG15 \\
      Xe & 031PAW \\
      Xe & 031US \\      
      \hline
    \end{tabular}
    \caption{List of the only pseudopotentials having ghost states in the conduction bands up to 10 eV above the Fermi level, for all libraries tested. Obviously, none of these is included in the SSSP libraries. The latest release of the SG15 library (SG15-1.1) has no ghost states.}\label{ghost_states}
  \end{center}
\end{table*}
}

\begin{table*}[!hbtp]
  \begin{center}
    \begin{tabular}{|c|c|c|c|c|c|c|}
     \hline
    \begin{normalsize} Element \end{normalsize} & \multicolumn{3}{c|}{\begin{normalsize} SSSP efficiency \end{normalsize}} &  
     \multicolumn{3}{c|}{\begin{normalsize} SSSP precision \end{normalsize}}\\
     \cline{2-4}\cline{5-7}
   (1-38)  & Pseudopotential & Cutoff & Dual &  Pseudopotential & Cutoff & Dual \\ 
      \hline
H & 100US & 60.0 & 8.0 & SG15 & 80.0 & 4.0 \\
He & SG15 & 50.0 & 4.0 & SG15 & 55.0 & 4.0 \\
Li & GBRV-1.4 & 40.0 & 8.0 & GBRV-1.4 & 40.0 & 8.0 \\
Be & GBRV-1.4 & 40.0 & 8.0 & SG15 & 55.0 & 4.0 \\
B & GBRV-1.4 & 35.0 & 8.0 & GBRV-1.2 & 55.0 & 8.0 \\
C & 100PAW & 45.0 & 8.0 & 100PAW & 45.0 & 8.0 \\
N & THEOS & 60.0 & 8.0 & Dojo & 80.0 & 4.0 \\
O & 031PAW & 50.0 & 8.0 & 031PAW & 75.0 & 8.0 \\
F & GBRV-1.4 & 45.0 & 8.0 & Dojo & 90.0 & 4.0 \\
Ne & SG15 & 50.0 & 4.0 & SG15 & 50.0 & 4.0 \\
Na & GBRV-1.5 & 40.0 & 8.0 & SG15 & 100.0 & 4.0 \\
Mg & 031PAW & 30.0 & 8.0 & GBRV-1.4 & 45.0 & 8.0 \\
Al & 100PAW & 30.0 & 8.0 & 100PAW & 30.0 & 8.0 \\
Si & 100US & 30.0 & 8.0 & 100US & 30.0 & 8.0 \\
P & 100US & 30.0 & 8.0 & 100US & 30.0 & 8.0 \\
S & GBRV-1.4 & 35.0 & 8.0 & GBRV-1.4 & 35.0 & 8.0 \\
Cl & GBRV-1.4 & 40.0 & 8.0 & 100US & 100.0 & 8.0 \\
Ar & SG15-1.1 & 60.0 & 4.0 & SG15-1.1 & 120.0 & 4.0 \\
K & 100PAW & 60.0 & 8.0 & 100PAW & 60.0 & 8.0 \\
Ca & GBRV-1.2 & 30.0 & 8.0 & GBRV-1.2 & 30.0 & 8.0 \\
Sc & SG15 & 40.0 & 4.0 & 031PAW & 90.0 & 8.0 \\
Ti & GBRV-1.4 & 35.0 & 8.0 & GBRV-1.4 & 40.0 & 8.0 \\
V & GBRV-1.4 & 35.0 & 8.0 & GBRV-1.4 & 40.0 & 8.0 \\
Cr & GBRV-1.5 & 40.0 & 8.0 & GBRV-1.5 & 40.0 & 8.0 \\
Mn & GBRV-1.5 & 65.0 & 12.0 & GBRV-1.5 & 90.0 & 12.0 \\
Fe & 031PAW & 90.0 & 12.0 & 031PAW & 90.0 & 12.0 \\
Co & GBRV-1.2 & 45.0 & 8.0 & GBRV-1.2 & 90.0 & 12.0 \\
Ni & GBRV-1.4 & 45.0 & 8.0 & GBRV-1.4 & 50.0 & 8.0 \\
Cu & GBRV-1.2 & 55.0 & 8.0 & SG15 & 90.0 & 4.0 \\
Zn & GBRV-1.2 & 40.0 & 8.0 & GBRV-1.2 & 90.0 & 8.0 \\
Ga & 100PAW & 70.0 & 8.0 & 100PAW & 90.0 & 8.0 \\
Ge & GBRV-1.4 & 40.0 & 8.0 & GBRV-1.4 & 45.0 & 8.0 \\
As & 031US & 35.0 & 8.0 & 031US & 35.0 & 8.0 \\
Se & GBRV-1.2 & 30.0 & 8.0 & GBRV-1.2 & 30.0 & 8.0 \\
Br & GBRV-1.4 & 30.0 & 8.0 & GBRV-1.4 & 90.0 & 8.0 \\
Kr & SG15 & 45.0 & 4.0 & SG15 & 50.0 & 4.0 \\
Rb & SG15 & 30.0 & 4.0 & SG15 & 30.0 & 4.0 \\
Sr & GBRV-1.2 & 30.0 & 8.0 & GBRV-1.2 & 40.0 & 8.0 \\\hline
    \end{tabular}
    \caption{SSSP efficiency and precision libraries version 1.1 (part I). The suggested wavefunction cutoffs (in Ry) and duals are also reported.}\label{sssp_pseudo}
  \end{center}
\end{table*}

\begin{table*}[!hbtp]
  \begin{center}
    \begin{tabular}{|c|c|c|c|c|c|c|}
     \hline
    \begin{normalsize} Element \end{normalsize} & \multicolumn{3}{c|}{\begin{normalsize} SSSP efficiency \end{normalsize}} &  
     \multicolumn{3}{c|}{\begin{normalsize} SSSP precision \end{normalsize}}\\
     \cline{2-4}\cline{5-7}
   (39-85)  & Pseudopotential & Cutoff & Dual &  Pseudopotential & Cutoff & Dual \\ 
      \hline
Y & GBRV-1.2 & 35.0 & 8.0 & GBRV-1.2 & 35.0 & 8.0 \\
Zr & GBRV-1.2 & 30.0 & 8.0 & GBRV-1.2 & 30.0 & 8.0 \\
Nb & 031PAW & 40.0 & 8.0 & 031PAW & 40.0 & 8.0 \\
Mo & SG15 & 35.0 & 4.0 & SG15 & 35.0 & 4.0 \\
Tc & SG15 & 30.0 & 4.0 & SG15 & 40.0 & 4.0 \\
Ru & SG15 & 35.0 & 4.0 & SG15 & 35.0 & 4.0 \\
Rh & SG15 & 35.0 & 4.0 & SG15 & 55.0 & 4.0 \\
Pd & SG15 & 45.0 & 4.0 & SG15 & 50.0 & 4.0 \\
Ag & SG15 & 50.0 & 4.0 & SG15 & 55.0 & 4.0 \\
Cd & 031US & 60.0 & 8.0 & 031US & 90.0 & 8.0 \\
In & 031US & 50.0 & 8.0 & 031US & 50.0 & 8.0 \\
Sn & GBRV-1.2 & 60.0 & 8.0 & GBRV-1.2 & 70.0 & 8.0 \\
Sb & GBRV-1.4 & 40.0 & 8.0 & GBRV-1.4 & 55.0 & 8.0 \\
Te & GBRV-1.2 & 30.0 & 8.0 & GBRV-1.2 & 30.0 & 8.0 \\
I & 031PAW & 35.0 & 8.0 & 031PAW & 45.0 & 8.0 \\
Xe & SG15-1.1 & 60.0 & 4.0 & SG15-1.1 & 80.0 & 4.0 \\
Cs & GBRV-1.2 & 30.0 & 8.0 & GBRV-1.2 & 30.0 & 8.0 \\
Ba & 100PAW & 30.0 & 8.0 & 100PAW & 35.0 & 8.0 \\
La & Wentzcovitch & 40.0 & 8.0 & Wentzcovitch & 40.0 & 8.0 \\
Ce & Wentzcovitch & 40.0 & 8.0 & Wentzcovitch & 50.0 & 8.0 \\
Pr & Wentzcovitch & 40.0 & 8.0 & Wentzcovitch & 40.0 & 8.0 \\
Nd & Wentzcovitch & 40.0 & 8.0 & Wentzcovitch & 40.0 & 8.0 \\
Pm & Wentzcovitch & 40.0 & 8.0 & Wentzcovitch & 40.0 & 8.0 \\
Sm & Wentzcovitch & 40.0 & 8.0 & Wentzcovitch & 40.0 & 8.0 \\
Eu & Wentzcovitch & 40.0 & 8.0 & Wentzcovitch & 40.0 & 8.0 \\
Gd & Wentzcovitch & 40.0 & 8.0 & Wentzcovitch & 40.0 & 8.0 \\
Tb & Wentzcovitch & 40.0 & 8.0 & Wentzcovitch & 40.0 & 8.0 \\
Dy & Wentzcovitch & 40.0 & 8.0 & Wentzcovitch & 40.0 & 8.0 \\
Ho & Wentzcovitch & 40.0 & 8.0 & Wentzcovitch & 40.0 & 8.0 \\
Er & Wentzcovitch & 40.0 & 8.0 & Wentzcovitch & 40.0 & 8.0 \\
Tm & Wentzcovitch & 40.0 & 8.0 & Wentzcovitch & 40.0 & 8.0 \\
Yb & Wentzcovitch & 40.0 & 8.0 & Wentzcovitch & 40.0 & 8.0 \\
Lu & Wentzcovitch & 45.0 & 8.0 & Wentzcovitch & 45.0 & 8.0 \\
Hf & Dojo & 50.0 & 4.0 & Dojo & 55.0 & 4.0 \\
Ta & GBRV-1.2 & 45.0 & 8.0 & GBRV-1.2 & 50.0 & 8.0 \\
W & GBRV-1.2 & 30.0 & 8.0 & GBRV-1.2 & 50.0 & 8.0 \\
Re & GBRV-1.2 & 30.0 & 8.0 & GBRV-1.2 & 30.0 & 8.0 \\
Os & GBRV-1.2 & 40.0 & 8.0 & GBRV-1.2 & 40.0 & 8.0 \\
Ir & GBRV-1.2 & 55.0 & 8.0 & GBRV-1.2 & 65.0 & 8.0 \\
Pt & GBRV-1.4 & 35.0 & 8.0 & 100US & 100.0 & 8.0 \\
Au & SG15 & 45.0 & 4.0 & SG15 & 50.0 & 4.0 \\
Hg & SG15 & 50.0 & 4.0 & SG15 & 55.0 & 4.0 \\
Tl & GBRV-1.2 & 50.0 & 8.0 & GBRV-1.2 & 70.0 & 8.0 \\
Pb & 031PAW & 40.0 & 8.0 & 031PAW & 45.0 & 8.0 \\
Bi & GBRV-1.2 & 45.0 & 8.0 & GBRV-1.2 & 50.0 & 8.0 \\
Po & 100US & 75.0 & 8.0 & 100US & 80.0 & 8.0 \\
Rn & 100PAW & 120.0 & 8.0 & 100PAW & 200.0 & 8.0 \\
\hline
    \end{tabular}
    \caption{SSSP efficiency and precision libraries version 1.1 (part II). The suggested wavefunction cutoffs (in Ry) and duals are also reported.}\label{sssp_pseudo1}
  \end{center}
\end{table*}

\end{document}